\documentclass[onecolumn]{aa}
\usepackage[dvips]{graphicx}
\usepackage{txfonts}

\begin{document}
   \title{Survey of long-term variability of stars}

   \subtitle{I. Reliability of magnitudes in old star catalogues}

   \author{T. Fujiwara,
          \inst{1}
          H. Yamaoka\inst{2}
          \and
          S. J. Miyoshi\inst{1}
          }

   \offprints{T. Fujiwara}

   \institute{Department of Physics, Graduate School of Science, 
              Kyoto Sangyo University,
              Kamigamo Motoyama, Kita-ku, Kyoto 603-8555, Japan \\
              \email{tomochan@fuujin.kyoto-su.ac.jp} \\
              \email{sjm@gold.ocn.ne.jp}
         \and
              Department of Physics, Faculty of Science, Kyushu University,
              4-2-1, Ropponmatsu, Chuo-ku, Fukuoka 810-8560, Japan \\
              \email{yamaoka@rc.kyushu-u.ac.jp}              
              }
\date{Received 2 June 2003 / Accepted 30 September 2003}

   \abstract{
   The comparison of visual magnitudes of stars compiled in old 
   catalogues is expected to yield information about their long-term 
   magnitude variations. 
   In seven old catalogues whose historical data have been 
   intensively compared, 2123 sampled stars have been studied, 
   disregarding stars that we could not identify, double stars which 
   could be misidentified, or stars observed under poor conditions, 
   and known variable stars with large amplitude discrepancies. 
   The independence of stellar magnitude catalogues 
   is demonstrated by comparing seven old studies to 
   each other, suggesting that the magnitude estimates in each 
   catalogue reflect the brightness at each observational period. 
   Furthermore, by comparing them with a modern star catalogue, the
   magnitude differences show a Gaussian distribution. 
   Therefore, if they are sufficiently larger than the deduced
   standard deviations, the magnitude variations between the catalogues 
   can be considered real.
   Thus, the stellar magnitudes compiled in old studies 
   can be used as scientific data within the average 
   intrinsic uncertainty. 
   These seven old catalogues can be used as data for the 
   survey of the long-term variability of stars.
         
   
   }

   \maketitle
%

\section{Introduction}

   Approximately 100 years have passed since 
   the first systematic and accurate observations of fixed 
   stars were catalogued. 
   However it is still possible that there are stars with 
   variability on longer timescales than 100 years. 
   Our goal here is to study such long-term variable 
   stars and find the nature of their magnitude variations 
   through the survey of stellar magnitudes in old star catalogues.
   
   Long-term variabilities with large amplitudes can occur through 
   several mechanisms; an eclipsing binary is one possible example.
   Currently, the longest period of a known eclipsing binary is 27.1 
   years of $\varepsilon$ Aur, varying between 3.37 -- 3.91 magnitude. 
   This variance can be recognized by naked-eye observations.
   One could assume that the duration of the minimum should be longer 
   for a system with a longer orbital period, thus the variability 
   of such a system can be overlooked if the observation was made 
   within a short period.
   Essentially, it means that the photometric observations for some 
   catalogues may have been performed during an eclipse where 
   the recorded magnitudes could be at the minimum, while other 
   observations show the magnitudes out of an eclipse.

   Except within the eclipsing binaries, variabilities 
   with timescales longer than those of Mira-type pulsating 
   variables ($\sim$1 year) have rarely been observed. 
   The timescale of Mira-type variables is measured in the stellar 
   dynamic timescale, which cannot be longer than several years. 
   On the other hand, some variabilities with timescales 
   longer than 1 year may have been recorded. 
   The helium flash at the core of an intermediate mass star 
   which leads a star from the red giant branch to the horizontal 
   branch is one possible example. 
   Another example is a final helium shell flash and a thermal pulse 
   stage, as in FG Sge (Herbig and Boyarchuk \cite{herbig}) or V4334 
   Sgr =`Sakurai's Object' (Duerbeck et al. \cite{duerbeck}). 
   
   Other violent variables can be observed, such as S Dor-type 
   variables (P Cyg, $\eta$ Car) or ones with uncertain mechanisms 
   like V838 Mon. 
   One such variable is the widely recognized $\delta$ Sco, which 
   brightened unexpectedly from 2.3 mag to about 1.8 mag since July 
   2000. 
   Previously known as a stable normal B star, this star 
   is now classified as an eruptive irregular variable of the 
   $\gamma$ Cas-type. 
   This type of star is a rapidly rotating B III-IVe star with mass 
   outflow from its equatorial zone. 
   The formation of equatorial rings or disks is often accompanied by 
   temporary fading. 
   Light amplitudes may reach 1.5 mag in V. (Otero et al. \cite{otero}; 
   Fabregat et al. \cite{fabregat}). 
   The variability of $\delta$ Sco was not expected. 
   As in this star, there may be many magnitude variations 
   hitherto unknown. 
   We expect that our survey will reveal such astronomical phenomena.
     
   To study stellar magnitudes of earlier eras, we referred 
   to old astronomical catalogues. 
   The following seven catalogues have been selected as reliable:
        
   \begin{enumerate}
     \item {\it Almagest} (Ptolemy AD127--141)
     \item {\it Kit\={a}b \d{S}uwar al-Kaw\={a}kib} (al-\d{S}\={u}f\={\i} 986)
     \item {\it Ulugh Beg's Catalogue of stars} (1437)
     \item {\it Astronomiae Instauratae Progymnasmata} (Brahe 1602)
     \item {\it Uranometria} (Bayer 1603)
     \item {\it Historia Coelestis Britannica} (Flamsteed 1725)
     \item {\it Uranometria Nova} (Argelander 1843)
   \end{enumerate}

   Before we could use the above studies, we had to check their 
   reliability as scientific data.
   We analyzed this problem on the basis of a statistical test 
   of the distribution of magnitude differences taken from each 
   pair of these studies. 
   As an additional check on the reliability of the seven 
   historical catalogues, we also compared the data compiled 
   in them with modern data taken from the `{\it Sky Catalogue 2000.0}' 
   (Hirshfeld et al. \cite{hirshfeld}).
   

\section{Characteristics of old catalogues}
   
   `{\it Almagest}' was compiled by Ptolemy in the 2nd 
   century AD. 
   Intensive philological studies of `{\it Almagest}' were 
   conducted by Kunitzsch (\cite{kunitzsch1986}) and Toomer 
   (\cite{toomer}). 
   We used the star catalogues of these two works, which contain 
   1022 stars of 48 constellations compiled with their ecliptic 
   coordinates and visual magnitudes. 
   Ptolemy's own recorded observations range from AD 127 to 
   141 and his catalogue epoch is about AD 137. 

   `{\it Kit\={a}b \d{S}uwar al-Kaw\={a}kib}' was written in Arabic 
   in the 10th century by al-\d{S}\={u}f\={\i}. 
   The epoch of this star table is 964. 
   Since the manuscript was transcribed by hand, our most 
   serious concern was that a clerical error may have been made. 
   We examined many manuscripts and literature relevant to this 
   material (al-\d{S}\={u}f\={\i} (\cite{sufia} \cite{sufib}); 
   al-B\={\i}r\={u}n\={\i} (\cite{biruni}); Schjellerup 
   (\cite{schjellerup}); Kunitzsch (\cite{kunitzsch1989})), 
   and adopted relative magnitude data for all of the records. 
   If there was a discrepancy between catalogues, 
   we followed the studies of Kunitzsch (\cite{kunitzsch1989}). 
   
   Knobel (\cite{knobel}) revised `{\it Ulugh Beg's Catalogue 
   of Stars}' using all the contemporary Persian manuscripts 
   kept in Great Britain. 
   This catalogue also includes star coordinates and magnitudes, 
   as observed from Samarkand (epoch 1437).
   
   Tycho Brahe observed a supernova in Cassiopeia (Tycho's nova) 
   in 1572 and recorded it in two books (Brahe \cite{brahe1573}, 
   \cite{brahe1602}). 
   One of them, `{\it Astronomiae Instauratae Progymnasmata}' 
   was published after his death in 1602 and includes a star 
   catalogue. 
   The data in this catalogue are based on Tycho's own 
   observations and are highly precise, especially in the 
   determination of stellar positions (errors are within 1$\arcmin$).
   
   Bayer introduced a new method to name fixed stars 
   in `{\it Uranometria}' (Bayer \cite{bayer}). 
   He named each star, per constellation, with Greek or Roman 
   alphabet characters in order of magnitude. 
   Until then, the identification of stars was usually done 
   by numbers and means of elaborate descriptions: 
   for example,  $\alpha$ UMi was described as `the 
   star on the end of the tail of the Little Bear'. 
   Bayer's identifications clarified obscure descriptions 
   and has been used widely up to the present. 
   In this material, Bayer added 12 southern constellations 
   to Ptolemy's original 48. 
   It depicts the positions and magnitudes of about 1200 stars. 
          
   Flamsteed performed positional astronomy at Royal Greenwich 
   Observatory and made 20,000 observations of nearly 3,000 stars. 
   His observational data, compiled in 
   `{\it Historia Coelestis Britannica}' (Flamsteed \cite{flamsteed}) 
   was published after his death. 
   This record is spread out over three volumes of 
   which the first two include data on planetary movement. 
   His star catalogue, including stellar equatorial coordinates, 
   ecliptic coordinates and magnitudes is contained in 
   Tome (Volume) III. 
   The observation epochs are described in the catalogue; 
   the mean epoch is 1689.
   
   Argelander observed a few thousand stars with the naked eye. 
   His `{\it Uranometria Nova}' (Argelander \cite{argelander}) 
   records 3256 stars with equatorial coordinates and magnitudes 
   of fixed stars.
   
   In all of the star catalogues mentioned above, stellar 
   magnitudes were classified from 1st to 6th based on the 
   Hipparchus system. 
   In addition, for refinement, observers used plus or 
   minus signs to indicate `a little brighter' or `a little dimmer', 
   respectively. 
   To quantify these magnitude descriptions, we subtracted 
   or added 0.33 according to the plus or minus sign, respectively. 
   For example, we assigned 2.67 for $3+$ and 3.33 for $3-$.

\section{Data selection and analysis}
     
   When these catalogues were recorded, there was no concept 
   of zero or minus magnitude. 
   Therefore, the stars brighter than 1 mag were omitted. 
   In addition we omitted the stars that we could not identify. 
   For example, since Bayer recorded the six stars $\pi^{1}$, 
   $\pi^{2}$ $\cdots$ $\pi^{6}$ Ori all together and described them 
   as $\pi$ Ori, we could not assign them individual magnitudes. 
   The constellation `Argo' was divided into 4 constellations 
   (Puppis, Pyxis, Vela, Carina) in the 18th century by Lacaille. 
   We could not identify the stars belonging to `Argo' in the old 
   star catalogues. 
   We omitted visual double stars and binaries (except 
   for spectroscopic binaries) whose apparent distance exceeds 
   1$\arcmin$ (limit of the resolving power of the naked eye) 
   and recorded as one single object. 
   For example, the apparent distance between $\alpha^{1}$ Cap 
   and $\alpha^{2}$ Cap is 7$\arcmin$ and magnitude data recorded 
   as `$\alpha$ Cap' were rejected. 
   For close stars (separated by less than 1$\arcmin$), we used the 
   compiled magnitude from the old material. 
   For the present-day magnitudes of close multiple 
   stars, we used the combined magnitudes of component stars taken 
   from the `{\it Sky Catalogue 2000.0}'. 
   Known variables with amplitude larger than 0.5 magnitude 
   ($o$ Cet (Mira), $\beta$ Per (Algol), $\delta$ Cep, etc.) were 
   omitted.
   
   The catalogues we used were recorded or compiled by different 
   people at different places and times. 
   Therefore, it might be possible that the listed magnitudes show 
   discrepancies only because of different observational conditions. 
   To find these discrepancies and to correct them, we compared the 
   mean magnitude averaged over all stars listed in each study to the 
   mean magnitude of corresponding stars listed in the 
   `{\it Sky Catalogue 2000.0}'. 
   The mean magnitudes and discrepancies thus obtained are presented 
   in Table~\ref{Reduction}. 
   The catalogue $ID$ (listed in Sect.1) is found in Column 1, 
   the observational (usually not published) epoch of each 
   catalogue is given in Column 2, the total number of stars 
   in each catalogue $N_{total}$ is shown in Column 3, 
   the number of selected stars {\it N} is listed in Column 4, 
   the ratio of selected star $N/N_{total}$ is given in Column 5, 
   mean magnitude of the catalogue $\bar{m}$ is listed in Column 6, 
   present-day mean magnitude $\bar{m}_{2000}$ is given in Column 7, 
   and $\bar{m}$ - $\bar{m}_{2000}$ is shown in Column 8. 
   Most datasets were obtained over extended periods. 
   However, these are much shorter than the epoch 
   differences between the catalogues. 
   We therefore neglected errors of several years and adopted 
   probable epochs. 
   
   \begin{table}
      \caption[]{Mean magnitudes and discrepancies}
         \label{Reduction}
         $$
         \begin{array}{c@{\quad}r@{\quad}r@{\quad}r@{\quad}r@{\quad}l
         @{\quad}l@{\quad}r}
            \hline
            \noalign{\smallskip}
            ID & epoch & N_{total} & N & N/N_{total}(\%) & 
            \bar{m} & \bar{m}_{2000} & \bar{m} - \bar{m}_{2000} \\
            \noalign{\smallskip}
            \hline
            \noalign{\smallskip}
            1 & 137 & 1022 & 910 & 89 & 3.98 & 4.06 & -0.08 \\   
            2 & 964 & 1025 & 911 & 89 & 4.16 & 4.07 & 0.09 \\
            3 & 1437 & 1018 & 889 & 87 & 4.16 & 4.06 & 0.10 \\
            4 & 1572 & 777 & 658 & 85 & 4.27 & 4.08 & 0.19 \\
            5 & 1603 & \sim 1200 & 949 & \sim79 & 4.42 & 4.26 & 0.16 \\
            6 & 1689 & \sim 3000 & 1003 & \sim33 & 4.61 & 4.36 & 0.25 \\
            7 & 1843 & 3256 & 1946 & 60 & 5.03 & 4.81 & 0.22 \\                     

            \noalign{\smallskip}
            \hline
         \end{array} 
         $$
   \end{table}

   According to the catalogues, the number of selected stars 
   differed greatly. 
   For example, from the oldest catalogue, `{\it Almagest}', 
   out of 1022 stars, we use 910 (89\%), however only 1946 
   stars out of 3256 (60\%) are taken from the most recent 
   and probably most reliable list (Argelander). 
   We accepted only 30\% from the catalogue of Flamsteed. 
   In the case of these two catalogues there were special 
   reasons for the high rejection percentage. 
   In Flamsteed's catalogue, there were many stars 
   without identification marks (Bayer names or Flamsteed's 
   numbers) which were not selected. 
   Flamsteed's numbers were not found in `{\it Historia Coelestis 
   Britannica}' which we could consult at the Paris Observatory 
   but in another book. 
   As for the catalogue of Argelander, the stars identified by 
   neither Bayer names nor Flamsteed's numbers were not sampled. 
   We could not associate the other identification marks with 
   currently known ones.
   Therefore, we used the stars marked with common identifications.
   
\section{Results and Discussion}
  \subsection{Independence of catalogues}
  To investigate whether these catalogues were based on 
  individual observations or copied from predecessors, we 
  compared these seven catalogues to each other. 
  If material was copied from predecessors, their magnitude data 
  would be identical and the distribution of stellar 
  magnitude differences would have a very small standard deviation. 
  If the standard deviation is large, we could assume that the 
  magnitude data was observed independently. 
  The distribution of the differences of stellar magnitudes between 
  each pair of studies is shown in Fig.~\ref{Distribution1}.
  
   \begin{figure}[p]
   \centering
    \begin{tabular}{ccc}
      \resizebox{60mm}{!}{\includegraphics{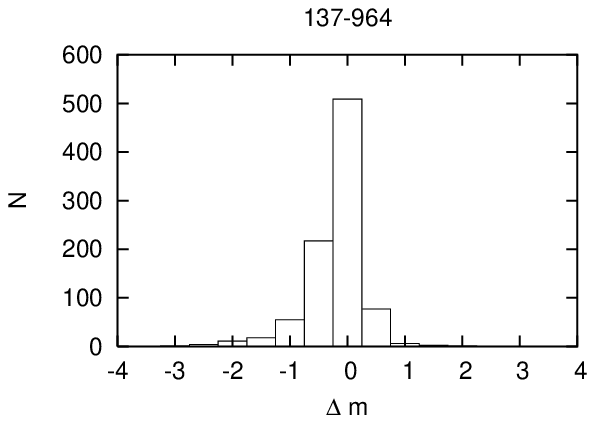}} &
      \resizebox{60mm}{!}{\includegraphics{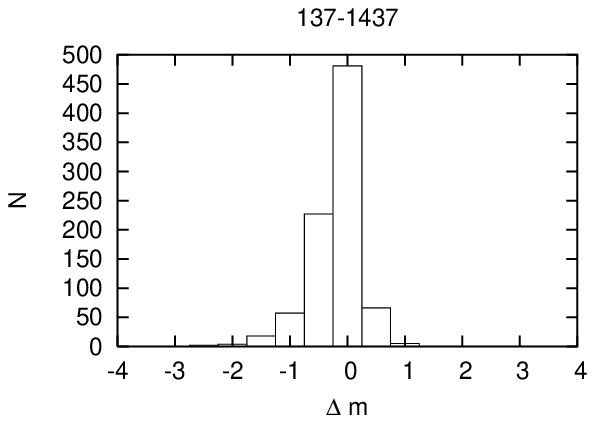}} &
      \resizebox{60mm}{!}{\includegraphics{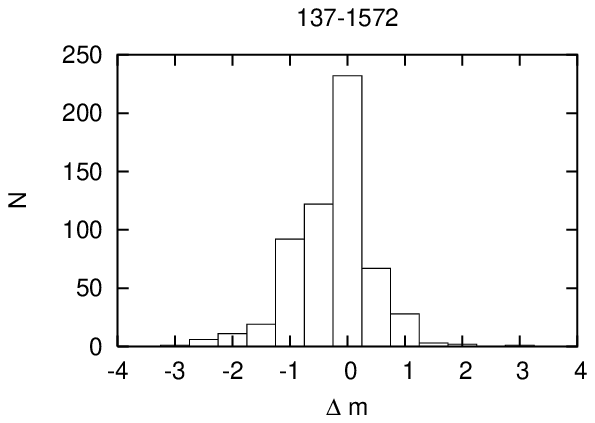}} \\
      \resizebox{60mm}{!}{\includegraphics{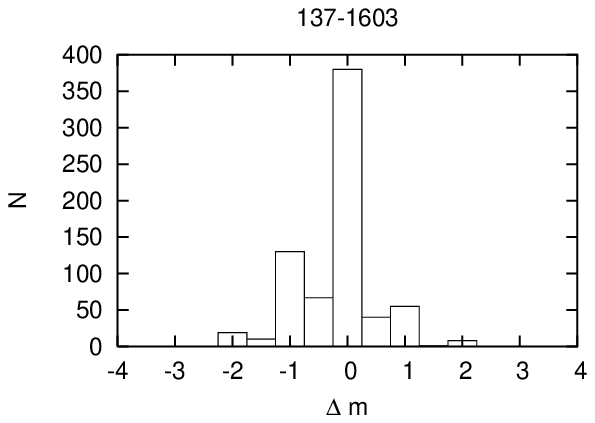}} &
      \resizebox{60mm}{!}{\includegraphics{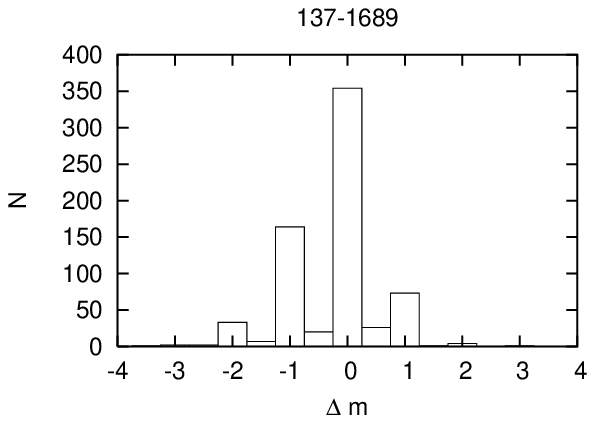}} &
      \resizebox{60mm}{!}{\includegraphics{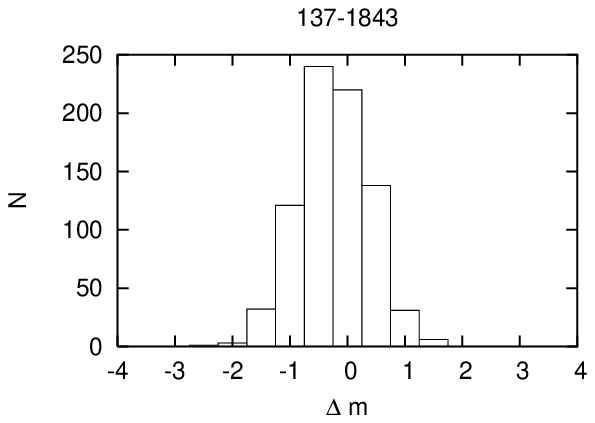}} \\
      \resizebox{60mm}{!}{\includegraphics{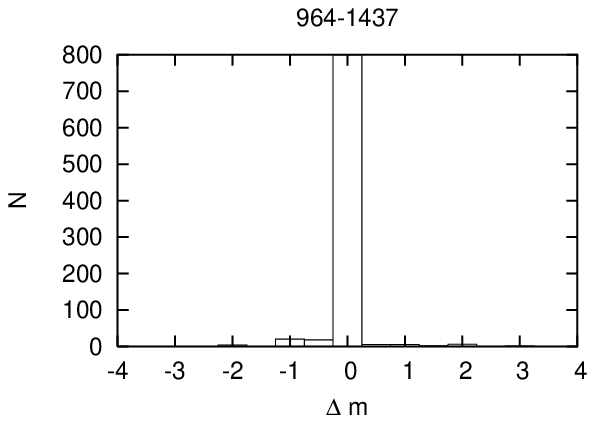}} &
      \resizebox{60mm}{!}{\includegraphics{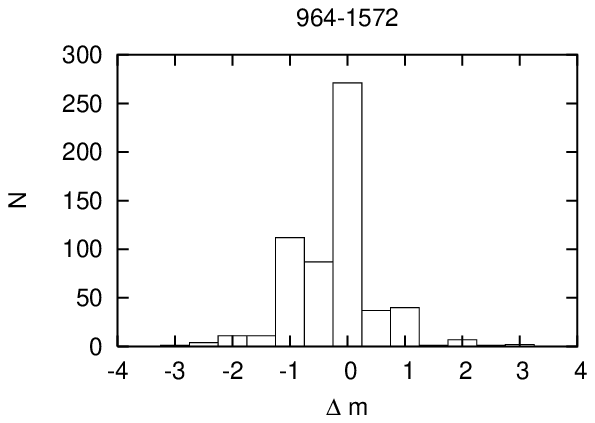}} &
      \resizebox{60mm}{!}{\includegraphics{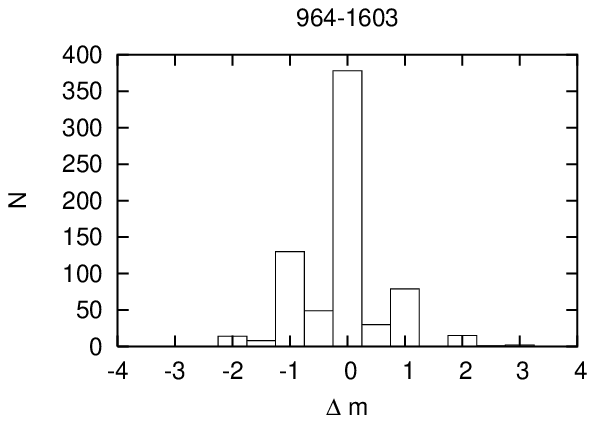}} \\
      \resizebox{60mm}{!}{\includegraphics{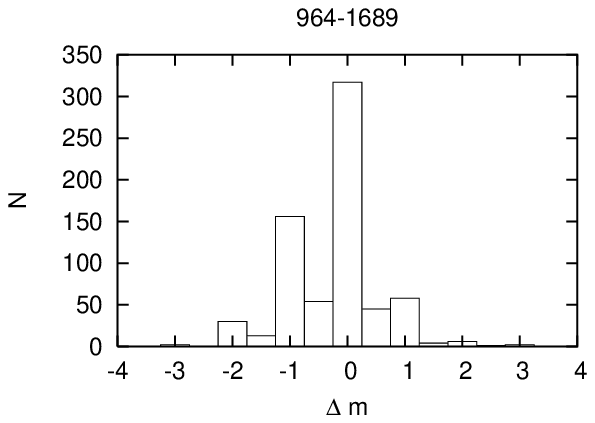}} &
      \resizebox{60mm}{!}{\includegraphics{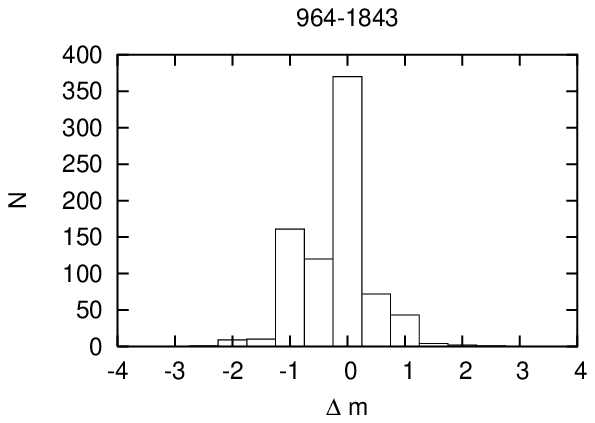}} &
      \resizebox{60mm}{!}{\includegraphics{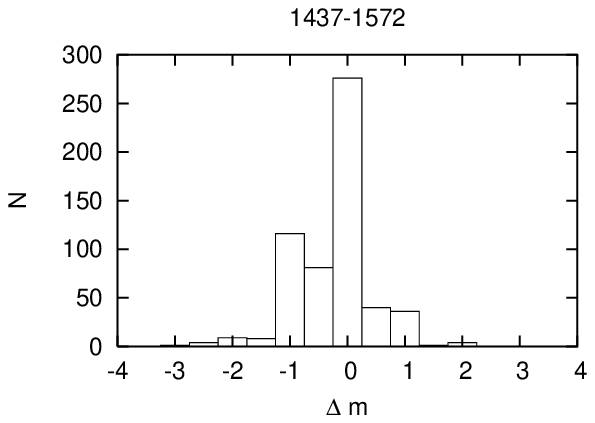}} \\
      \resizebox{60mm}{!}{\includegraphics{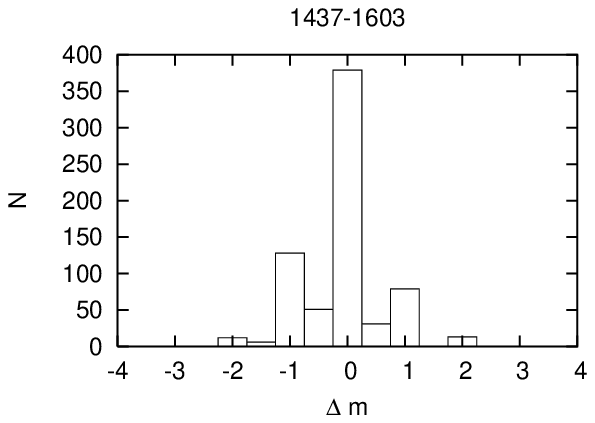}} &
      \resizebox{60mm}{!}{\includegraphics{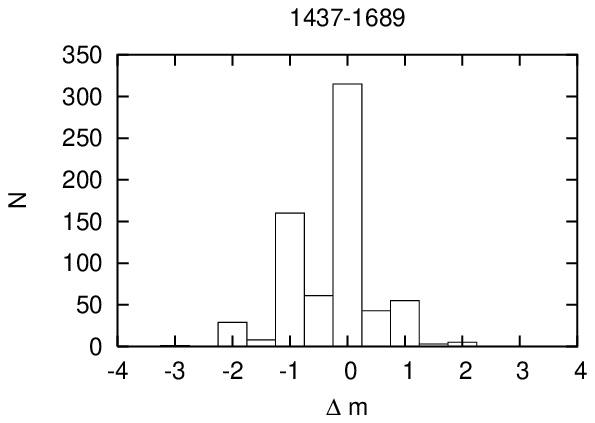}} &
      \resizebox{60mm}{!}{\includegraphics{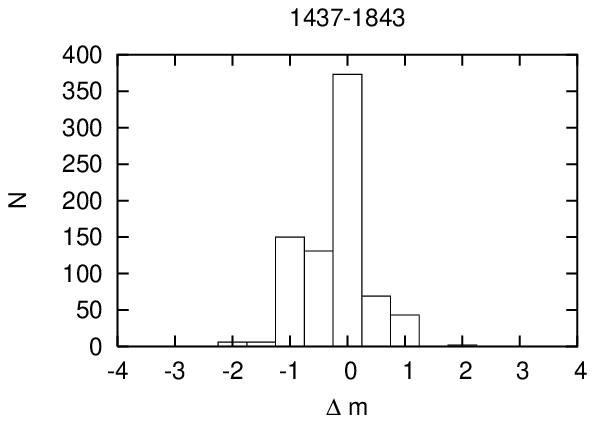}} \\
    \end{tabular}

      \caption{Differences of stellar magnitude between two old catalogues}
         \label{Distribution1}
   \end{figure}
%

\addtocounter{figure}{-1}
   \begin{figure}[p]
   \centering
    \begin{tabular}{ccc}
      \resizebox{60mm}{!}{\includegraphics{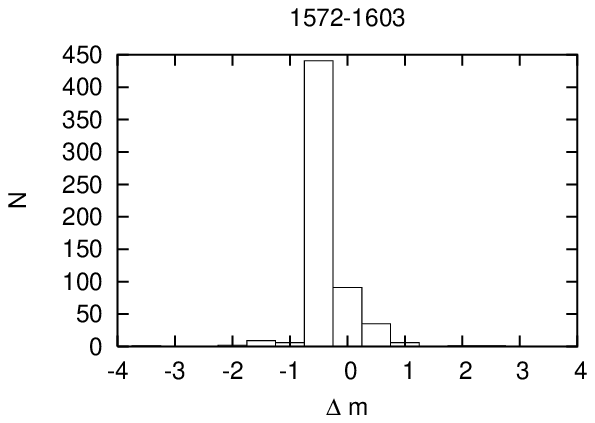}} &
      \resizebox{60mm}{!}{\includegraphics{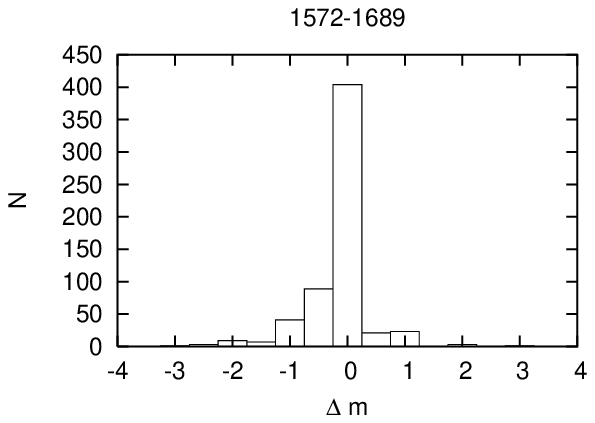}} &
      \resizebox{60mm}{!}{\includegraphics{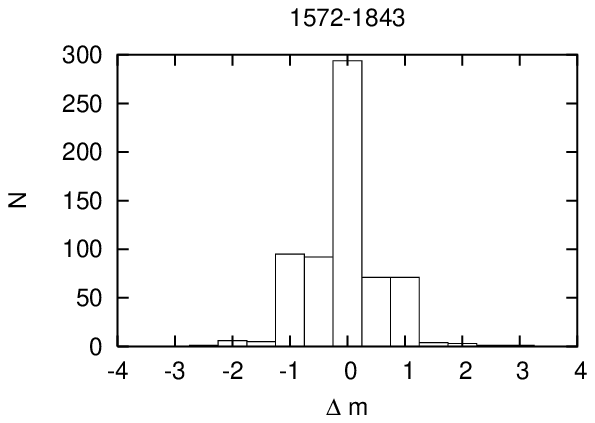}} \\
      \resizebox{60mm}{!}{\includegraphics{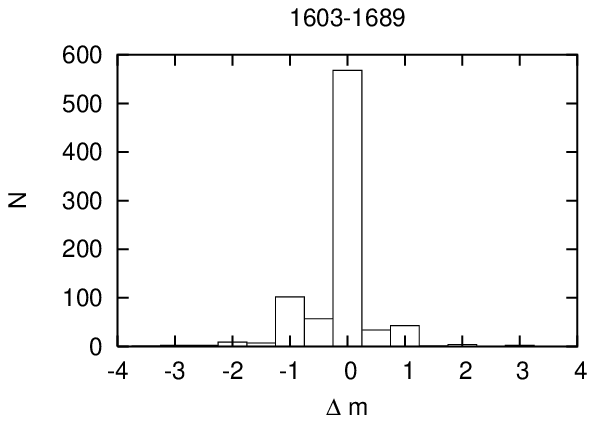}} &
      \resizebox{60mm}{!}{\includegraphics{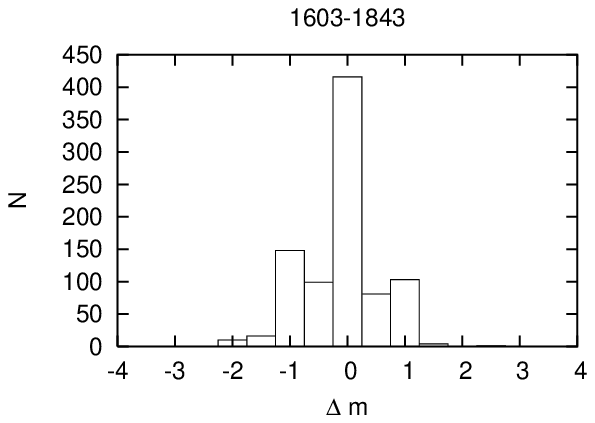}} &
      \resizebox{60mm}{!}{\includegraphics{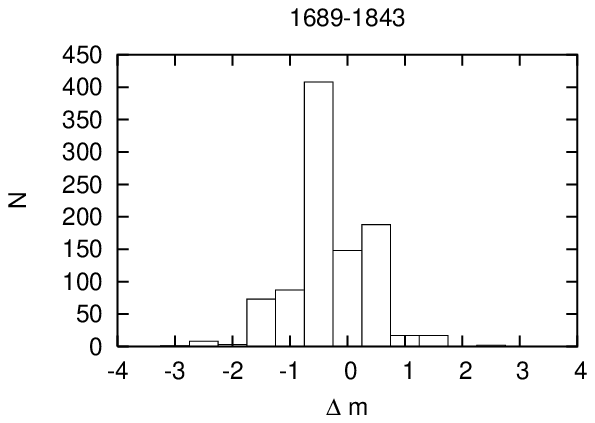}} \\
    \end{tabular}
      \caption{continued.}
   \end{figure}
%
  
  The standard deviations $\sigma$ of these distributions and the 
  numbers of sampled stars $N$ are given in 
  Table~\ref{DistributionTable1}. 
  To clarify the value of the standard deviations, we set 
  the mean of the magnitude difference for a pair of catalogues to 
  zero by adding a small (up to 0.1 mag) constant for each pair. 
  
  With the exception of the correlation between  
  `{\it Kit\={a}b \d{S}uwar al-Kaw\={a}kib}' and Ulugh Beg's catalogue, 
  the standard deviations $\sigma$ range from 0.40 to 0.78 mag 
  for all catalogue comparisons. 
  These values are much larger than expected for non-independent 
  records. 
  Therefore, the stellar magnitudes listed in these catalogues 
  are considered to have been observed independently. 

  The correlation between 
  `{\it Kit\={a}b \d{S}uwar al-Kaw\={a}kib} (964)' 
  and `{\it Ulugh Beg's Catalogue of stars} (1437)' is very close 
  with little to no deviation. 
  In Fig.\ref{Distribution1}, one can see the large peak in the 
  distribution difference graph comparing these two sets.
  
  Despite a span of over 450 years, the dispersion $\sigma$ is 
  much smaller than in the other correlations. 
  However, if most of Ulugh Beg's catalogue was copied from 
  `{\it Kit\={a}b \d{S}uwar al-Kaw\={a}kib}', the dispersion 
  should be close to 0. 
  The standard deviation of 0.29 mag indicates that Ulugh Beg's 
  catalogue is not a complete copy, but gives strong reasons 
  to suspect that the two catalogues are not fully independent 
  either. 

   \begin{table}
      \caption[]{Standard deviations $\sigma$ between two old catalogues 
      and numbers of sampled stars $N$}
         \label{DistributionTable1}
         $$
         \begin{array}{c@{\quad}c@{\quad}c@{\quad}c@{\quad}c@{\quad}c@
         {\quad}c@{\quad}c@{\quad}c@{\quad}c}
            \hline
            \noalign{\smallskip}
            ID & epoch & & 137 & 964 & 1437 & 1572 & 1603 & 1689 & 1843 \\
            & & & \multicolumn{7}{c}{\sigma}\\
            \noalign{\smallskip}
            \hline
            \noalign{\smallskip}
            1 & 137 &  & --- & 0.47 & 0.41 & 0.69 & 0.67 & 0.77 & 0.62 \\   
            2 & 964 &  & 901 & --- & 0.29 & 0.72 & 0.72 & 0.78 & 0.60 \\
            3 & 1437 &  & 860 & 861 & --- & 0.66 & 0.68 & 0.72 & 0.55 \\
            4 & 1572 & N & 584 & 585 & 575 & --- & 0.40 & 0.54 & 0.65 \\
            5 & 1603 &  & 709 & 706 & 699 & 593 & --- & 0.59 & 0.65 \\
            6 & 1689 &  & 688 & 688 & 680 & 602 & 832 & --- & 0.67 \\
            7 & 1843 &  & 792 & 793 & 780 & 644 & 878 & 952 &  ---   \\              
            \noalign{\smallskip}
            \hline
         \end{array} 
         $$
   \end{table}

\subsection{Consistency of magnitudes}

  In addition to this comparison, we compared the stellar 
  magnitudes in these old surveys with those in the 
  `{\it Sky Catalogue 2000.0}' (Hirshfeld et al. 
  \cite{hirshfeld}). 
  The standard deviations $\sigma$ of these distributions are 
  shown in Table~\ref{DistributionTable2}, and the number of 
  sampled stars $N$ are given in Table~\ref{Reduction}.
     
\begin{figure}[p]
   \centering
    \begin{tabular}{ccc}
      \resizebox{60mm}{!}{\includegraphics{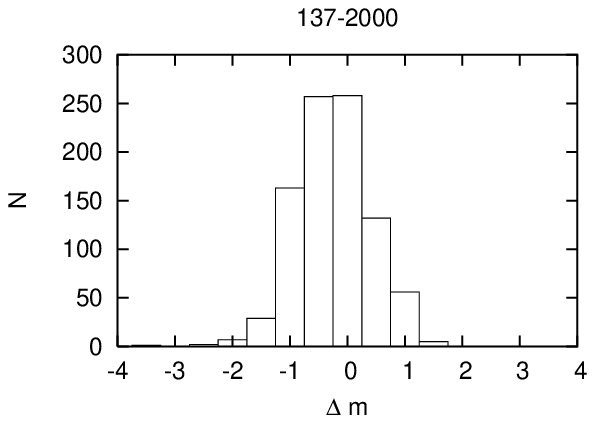}} &
      \resizebox{60mm}{!}{\includegraphics{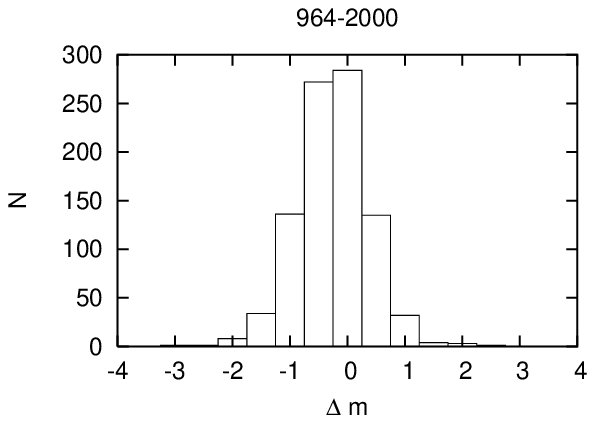}} &
      \resizebox{60mm}{!}{\includegraphics{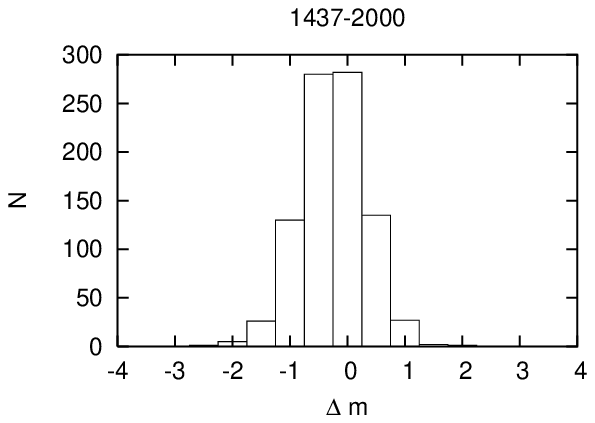}} \\
      \resizebox{60mm}{!}{\includegraphics{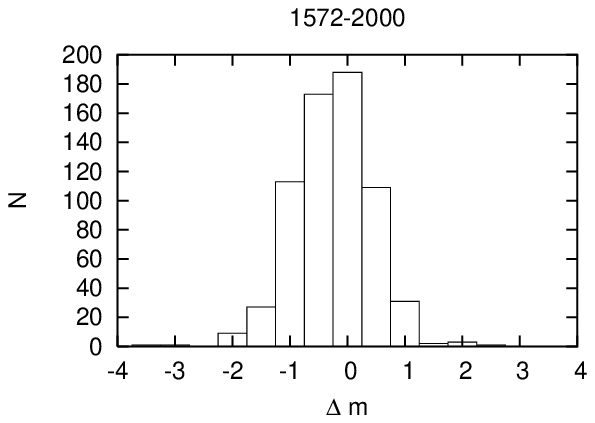}} &
      \resizebox{60mm}{!}{\includegraphics{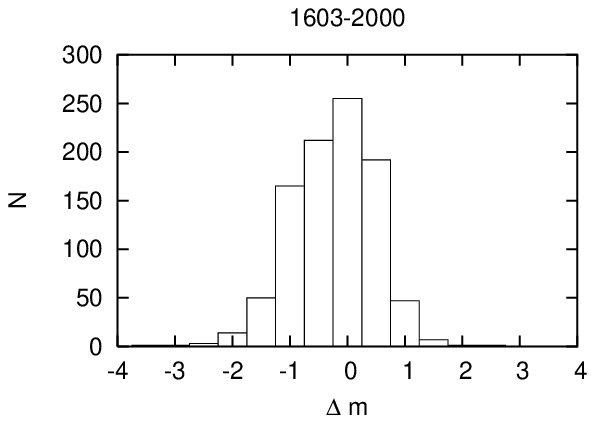}} &
      \resizebox{60mm}{!}{\includegraphics{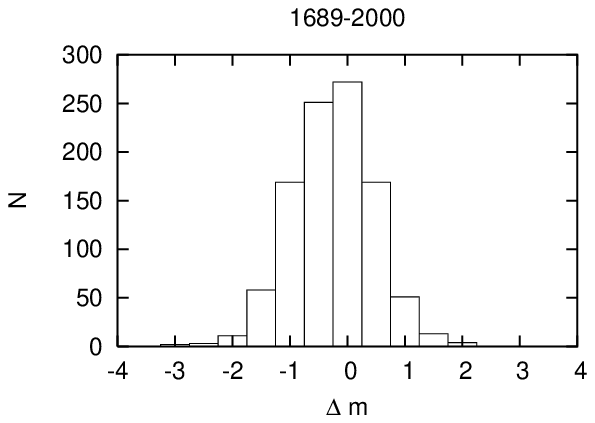}} \\
      \resizebox{60mm}{!}{\includegraphics{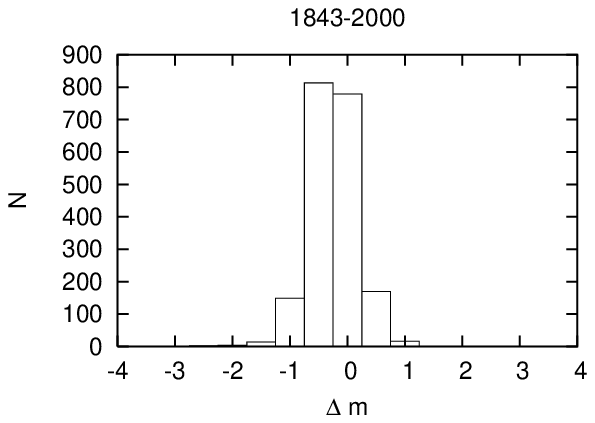}} \\
    \end{tabular}
      \caption{Differences of stellar magnitude between 
      old catalogues and `{\it Sky Catalogue 2000.0}'}
         \label{Distribution2}
   \end{figure}

  The standard deviations $\sigma$ range between 0.41 and 0.70. 
  Magnitude differences measured for any combination of old 
  catalogue and the `{\it Sky Catalogue 2000.0}' show 
  a Gaussian distribution (see Fig.~\ref{Distribution2}). 
  These facts demonstrate that their magnitude variations 
  are considered to be real when the variation is sufficiently 
  larger than the dispersion. 
  Therefore, we can use these star catalogues as scientific 
  data within an average intrinsic uncertainty of about 0.5 mag. 
  
  We show discrepancies between the mean magnitude 
  of the catalogue $\bar{m}$ and the present-day mean magnitude 
  $\bar{m}_{2000}$ in each catalogue as 
  $\bar{m}$ - $\bar{m}_{2000}$ in Table~\ref{Reduction}. 
  These discrepancies are much less than 0.5, negligibly 
  small for the discussion of dispersions. 
  At later epochs, the value shifts toward more positive 
  residuals and more stars were recorded. 
  We propose that these discrepancies were ascribable to 
  dimmer stars which were estimated imprecisely. 

%

   \begin{table}
      \caption[]{Standard deviations $\sigma$ between old catalogues 
      and `{\it Sky Catalogue 2000.0}'}
         \label{DistributionTable2}
         $$
         \begin{array}{c@{\quad}c@{\quad}c}

            \hline
            \noalign{\smallskip}
            ID & epoch & 2000 \\
            & & \sigma \\
            \noalign{\smallskip}
            \hline
            \noalign{\smallskip}
            1 & 137 & 0.64 \\   
            2 & 964 & 0.61 \\
            3 & 1437 & 0.56 \\
            4 & 1572 & 0.67 \\
            5 & 1603 & 0.70 \\
            6 & 1689 & 0.70 \\
            7 & 1843 & 0.41 \\                     
            \noalign{\smallskip}
            \hline
         \end{array} 
         $$
   \end{table}

\section{Conclusions}
   \begin{enumerate}
      \item The comparison of stellar magnitudes recorded in seven 
         old catalogues indicates that the magnitudes in most of 
         these catalogues were obtained from independent observations. 
         The only exception is `{\it Ulugh Beg's Catalogue of stars}', 
         which is probably not entirely independent of the earlier 
         `{\it Kit\={a}b \d{S}uwar al-Kaw\={a}kib}' list and 
         should therefore be given lower weight. 

      \item The magnitude variations found from the comparison of 
         stellar magnitudes recorded in seven old catalogues and one 
         modern star catalogue can be considered to be real.
               
      \item Magnitude differences between old catalogues and the 
         modern star catalogue also represent Gaussian 
         distributions, thereby supporting the above conclusions. 
         Essentially, the stellar magnitudes compiled in the old 
         studies we investigated here can be used as scientific data 
         within an average intrinsic uncertainty of about 0.5 mag. 
   \end{enumerate}

   Although it is beyond the scope of this paper, some of the 
   stars we could not identify and have omitted here might be 
   transient objects (nova, supernova or others) caught in an outburst.

\begin{acknowledgements}
      T. F. is grateful to Dr. P. Kunitzsch and professors in 
      Paris Observatory for their kind advice and valuable 
      information about the reliability of old catalogues.
            
      This research has made use of the Simbad database operated 
      at CDS, Strasbourg, France.
      
      This work is partly supported by Research Fellowships of the 
      Japan Society for the Promotion of Science for Young 
      Scientists (T.F.), and by a grant-in-aid [14740131 (H.Y.)] 
      from the Japanese Ministry of Education, Culture, Sports, 
      Science and Technology.
      
      The authors wish to thank the referee for useful advice 
      during the compilation of this paper. 

\end{acknowledgements}


\begin{thebibliography}{}

  \bibitem[1843]{argelander} Argelander, W. A. 1843,
      Uranometria Nova
      (Berlag von Simon Schropp und Comp., Berlin)

  \bibitem[1603]{bayer} Bayer, J. 1603,
      Uranometria, reprinted in 1987
      (British Library, England)

  \bibitem[1030]{biruni} al-B\={\i}r\={u}n\={\i} 1030,
      Al-Q\={a}n\={u}n al-Mas`\={u}d\={\i}, vol.3, reprinted in 1956 
      (D\={a}iratu'l-Ma`\={a}rif-il-`O\d{s}m\={a}nia, Hyderabad)

  \bibitem[1573]{brahe1573} Brahe, T. 1573,
      De nova et nullius aevi memoria prius visa stella, facs.ed. 
      in 1901
      (Copenhagen)
      
  \bibitem[1602]{brahe1602} Brahe, T. 1602,
      Astronomiae Instauratae Progymnasmata, reprinted in 1969
      (Impression Anastaltique, Bruxelles)

  \bibitem[1997]{duerbeck} Duerbeck, H. W. et al., 1997,
      \aj, {\bf 114}, 1657

  \bibitem[1725]{flamsteed} Flamsteed, J. 1725,
      Historia Coelestis Britannica, Tome III
      (London)

  \bibitem[2000]{fabregat} Fabregat, J., Reig, P., Otero, S. 2000,
      IAU Circular No. 7461

  \bibitem[1970--]{gillispie} Gillispie, C. C. 1970--,
      Dictionary of Scientific Biography
      (Scribner, New York)

  \bibitem[1996]{hearnshaw} Hearnshaw, J. B. 1996,
      The measurement of starlight
      (Cambridge U.P., Cambridge)
  
  \bibitem[1968]{herbig} Herbig, G. H. \& Boyarchuk, A. A. 1968,
      \apj, {\bf 153}, 397

  \bibitem[1991]{hirshfeld} Hirshfeld, A., Sinnott, R. W., \& Ochsenbein, F. 1991,
      Sky Catalogue 2000.0 vol.1, 2nd ed.
      (Cambridge U. P. \& Sky Publishing Corporation, Cambridge)

  \bibitem[1917]{knobel} Knobel, E. B. 1917,
      Ulugh Beg's Catalogue of stars (1437)
      (The Carnegie Institution of Washington, Washington)
      
  \bibitem[1989]{kunitzsch1989} Kunitzsch, P. 1989,
      The Arabs and the Stars
      (Variorum Reprints, London)

  \bibitem[1986-1991]{kunitzsch1986} Ptolem\"{a}us, C. 1986-1991,
      Der Sternkatalog des Almagest Die arabisch-mittelalterliche 
      Tradition Teil I-III (AD 2 c.), translated into German and edited 
      by Kunitzsch,P.
      (Otto Herrassowitz, Wiesbaden)

  \bibitem[2001]{otero} Otero, S., Fraser, B., Lloyd, C. 2001,
       International Bulletin on Variable Stars, 5026

  \bibitem[1874]{schjellerup} Schjellerup, H. C. F. C. 1874,
      Description des \'{e}toiles fixes, reprinted in 1987
      (Institut f\"{u}r Geschichte der Arabisch-Islamischen
       Wissenschaften, Frankfurt am Main)

  \bibitem[986a]{sufia} al-\d{S}\={u}f\={\i} 986,
      Kit\={a}b \d{S}uwar al-Kaw\={a}kib, edited in 1954 
      (D\={a}iratu'l-Ma`\={a}rif-il-`O\d{s}m\={a}nia, Hyderabad)

  \bibitem[986b]{sufib} al-\d{S}\={u}f\={\i} 986,
      Kit\={a}b \d{S}uwar al-Kaw\={a}kib (986), facs. ed. in 1986
      (Frankfurt am Main)

  \bibitem[1998]{toomer} Toomer, G. J. 1998,
      Ptolemy's Almagest (AD 2 c.), translated into English and annotated 
      by Toomer, G. J.
      (Princeton U. P., Princeton)

  \bibitem[1979]{warnerd} Warner, D. J. 1979,
      The Sky Explored: Celestial Cartography 1500-1800
      (Alan R. Liss, Inc. \& Theatrum Orbis Terrarum, B. V., New York \& Amsterdam)

  \bibitem[1986]{wernerh} Werner, H. \& Schmeidler, F. 1986,
      Synopsis der Nomenklatur der Fixsterne
      (Wissenschaftliche Verlagsgesellschaft mbH, Stuttgart)


\end{thebibliography}
\end{document}